\documentclass[12pt]{article} 

\usepackage{amsmath}        
\usepackage{epsfig,amssymb}
\usepackage{graphicx,psfrag}       

\usepackage[latin1]{inputenc}       

\newcommand{\re}[0]{\mathrm{Re}}
\newcommand{\im}[0]{\mathrm{Im}}
\newcommand{\chiup}{\raisebox{2pt}{$\chi$}}

\begin{document}

\title{\bf Experimental study of the
Fluctuation-Dissipation-Relation during an aging process}
\author{L. Bellon and  S. Ciliberto  \\
         Ecole Normale
Sup\'erieure de Lyon, Laboratoire de Physique ,\\
 C.N.R.S. UMR5672,  \\ 46, All\'ee d'Italie, 69364 Lyon Cedex
07,  France\\
        }
 \maketitle

 \begin{abstract}
 The validity of fluctuation dissipation relations in an aging system
  is studied in a colloidal glass during the
 transition from a fluid-like   to a solid-like state.
  The evolution of the rheological and electrical properties
 is analyzed in the range $1Hz - 40Hz$.
 It is found that at the beginning of the  transition the fluctuation
dissipation relation is strongly violated in electrical
measurements. The amplitude and the persistence time of this
violation are decreasing functions of frequency. At the lowest
frequencies of the measuring range it persists for times which are
about $5\%$ of the time needed to form the colloidal glass. This
phenomenology is quite close to the recent theoretical predictions
done for the violation of the fluctuation dissipation relation in
glassy systems. In contrast in the rheological measurements no
violation of the fluctuation dissipation relation is observed. The
reasons of this large difference between the electrical and
rheological measurements are discussed.
\end{abstract}

\section{Introduction}

Many physical systems in nature are not in thermodynamic
equilibrium because they present very slow relaxation processes. A
typical example of this phenomenon is the aging of glassy
materials: when they are quenched from above their glass
transition temperature $T_g$ to a temperature $T<T_g$, any
response function of these systems depends on the aging time $t_a$
spent at $T$. For example, the dielectric and elastic constants of
polymers continue to evolve several years after the quench
\cite{Struick}. Because of these slow relaxation processes, the
glass is out of equilibrium, and usual thermodynamics does not
apply. However, as this time evolution is slow, some concepts of
the classical approach may be useful for understanding the glass
aging properties. A widely studied question, is how the
temperature of these systems can be defined. One possible answer
comes from the study of the deviation to the Fluctuation
Dissipation Relation(FDR) in an out of equilibrium system (for a
review see ref. \cite{Mezard,Cugliandolo,Peliti}). In this letter
we show that this approach is relevant for the study of a
colloidal glass formation, where a strong violation of FDR is
measured. Implications of this observation go beyond the physics
interest. Indeed FDR is used as a tool to extract, from
fluctuations measurements, several properties in biological,
chemical and physical systems \cite{bio1,bio2,surface}. Our
results indicate that before extending this smart technique to
other systems, one has to carefully ensure that these systems are
in equilibrium.

In order to understand this new definition of temperature, we have
to recall the main consequences of FDR in a system which is in
thermodynamic equilibrium. We consider an observable $V$ of such a
system and its conjugate variables $q$ . The response function
$\chi_{Vq}(\omega)$, at frequency $\nu=\omega / 2 \pi$, describes
the variation $\delta V (\omega)$ of $V$ induced by a perturbation
$\delta q (\omega)$ of  $q$, that is $\chi_{Vq} (\omega)=\delta V
(\omega)/ \delta q(\omega)$. FDR relates the fluctuation spectral
density of $V$ to the response function $\chi_{Vq}$ and the
temperature T of the system:

\begin{equation}
S(\omega) = { 2 k_B \ T \over \pi \omega } {\it
Im}\left[\chi_{Vq}(\omega) \right] \label{FDR}
\end{equation}

where $S(\omega)=<|V(\omega)|^2>$ is the fluctuation spectral
density of $V$, $k_B$ is the Boltzmann constant, ${\it Im}\left[
\chi_{Vq}(\omega) \right]$ is the imaginary part of
$\chi_{Vq}(\omega)$. Textbook examples of FDR are Nyquist's
formula relating the voltage noise to the electrical resistance
and the Einstein's relation for Brownian motion relating the
particle diffusion coefficient  to the fluid viscosity
\cite{book}.

When the system is not in equilibrium FDR, that is eq.\ref{FDR},
may fail. For example violations, of about a factor of 2, of
eq.\ref{FDR} have been observed in the density fluctuations of
polymers in the glassy phase \cite{Wendorff}. The first to propose
that the study of the FDR violations are  relevant for glassy
systems was Sompolinsky \cite{Sompolinsky}. This idea, which was
generalized in the context of weak turbulence \cite{Shraiman}, has
been recently reconsidered by Cugliandolo and Kurchan
\cite{Kurchan} and successively tested in many analytical and
numerical models of
 glass dynamics
 \cite{Peliti},\cite{Parisi}-\cite{Berthier}.
Let us briefly recall the main and general findings of these
models. Because of the slow dependence on $t_a$ of the response
functions, it has been proposed that eq.\ref{FDR} can be used to
define an effective temperature of the system, specifically:

\begin{equation}
T_{eff} (t_a, \omega) = { S(t_a, \omega) \ \pi \omega \over  {\it
Im}\left[ \chi_{Vq}(t_a, \omega) \right] \ 2 k_B }
 \label{Teff}
\end{equation}

It is clear that if eq.\ref{FDR} is satisfied $T_{ef\!f}=T$,
otherwise $T_{ef\!f}$ turns out to be a decreasing function of
$t_a$ and $\omega$. The physical meaning of eq.\ref{Teff} is that
there is a time scale (for example $t_a$) which allows to separate
the fast processes from the slow ones. In other words the low
frequency modes relax towards the equilibrium value much slower
than the high frequency ones which rapidly relax to the
temperature of the thermal bath. Therefore it is conceivable that
the slow frequency modes keep memory of higher temperatures for a
long time and for this reason their temperature should be higher
than that of the high frequency ones. This striking behavior has
been observed in several numerical  models of aging
\cite{Peliti},\cite{Parisi}-\cite{Berthier}. Further analytical
and numerical studies of simple models show that eq.\ref{Teff} is
a good definition of temperature in the thermodynamic sense
\cite{Cugliandolo,Peliti}. In spite of the large amount of
theoretical studies there are only a few experiments which show a
violation of FDR in real materials \cite{Wendorff,Grigera}.
However these measurements are done at a single frequency and
there is no idea on how the temperature relaxes as function of
time and frequency. The experimental analysis of the dependence of
$T_{eff}(\omega,t_a)$ on $\omega$ and $t_a$ is very useful to
distinguish among different models of aging: FDR violations are
model dependent \cite{Peliti},\cite{Parisi}-\cite{Berthier}.

For these reasons we have experimentally studied the violation of
eq.\ref{FDR} during the colloidal glass formation in Laponite RD
\cite{Laponite}, a synthetic clay consisting of discoid charged
particles. It disperses rapidly in water  and solidifies even for
very low mass fraction. Physical properties of this preparation
evolves for a long time, even after the sol-gel transition, and
have shown many similarities with standard glass aging
\cite{Kroon,Bonn}. Recent experiments \cite{Bonn} have proved that
the structure function of Laponite at low concentration (less than
$3 \%$ mass fraction) is close to that of a glas.  As  in our
experiment the Laponite concentration is low, we  call the solid
like Laponite solution either a colloidal glass or simply a glass.

In our experiment we measure the time evolution of the Laponite
electrical and rheological  properties   during the colloidal
glass formation. The paper is organized as follows. In the next
section the electrical measurements are described. Rheological
measurements are discussed in section 3. Finally we conclude in
section 4.

\section{Electrical measurements}

\subsection{The experimental apparatus}

The Laponite solution is used as a conductive liquid between the
two golden coated electrodes of a cell (see fig.1). The Laponite
solution is prepared in a clean $\mathrm{N_2}$ atmosphere to avoid
$\mathrm{CO_2}$ and $\mathrm{O_2}$ contamination, which perturbs
the electrical measurements. Laponite particles are dissolved at a
concentration of $2.5 \%$ mass fraction in pure water under
vigorous stirring during $20 min$. To avoid the existence of any
initial structure in the sol, we pass the solution through a
 $1\mu m$ filter when filling our cell. This instant defines the origin of
the aging time $t_a$ (the filling of the cell takes roughly two
minutes, which can be considered the maximum inaccuracy of $t_a$).
The sample is then sealed so that no pollution or evaporation of
the solvent can occur. At this concentration, the  light
scattering experiments show that Laponite structure functions are
still evolving $500h$ after the preparation \cite{Kroon}.  We only
study the beginning of this glass formation process.

\begin{figure}[!ht ]
 \centerline{\epsfxsize=0.6\linewidth \epsffile{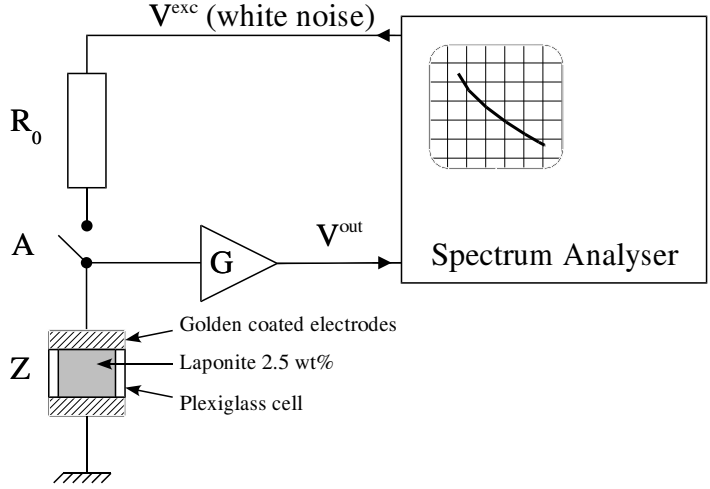}}
  \caption{{\bf Experimental set-up} The impedance under test $Z$ is a
cell (diameter $7cm$, thickness $3cm$) filled with a $2.5 wt\%$
Laponite sol. The electrodes of the cell are golden coated to
avoid oxidation. One of the two electrodes is grounded whereas the
other is connected to the entrance of a low noise  voltage
amplifier characterized by a voltage amplification $G$. With a
spectrum analyzer, we alternately record the frequency response
$F\!R(\omega)=<V^{out}/V^{exc}>$ (switch $A$ closed) and the
spectrum $S(\omega)=<|{V^{out}}|^2>$ (switch $A$ opened). The
input voltage $V^{exc}$ is a white noise excitation, thus from
$F\!R(\omega)$ we derive the impedance $Z(\omega)$ as a function
of $\omega$, that is $Z(\omega)=R_0 / (G / F\!R(\omega) - 1)$;
whereas from $S(\omega)$, we can estimate the voltage noise of
$Z$, specifically $S_Z(\omega) = [S(\omega)-S_a(\omega)]/G^2$
where $S_a(\omega)$ is the noise spectral density of the
amplifier} \label{measurement}
\end{figure}

The two electrodes of the cell are connected  to  our measurement
system, where we alternately record the cell electrical impedance
$Z(t_a,\omega)$ and the  voltage noise density $S_{Z}(t_a,\omega)$
(see Fig.\ref{measurement}). Taking into account that  in this
configuration ${\it Im }\left[ \chi_{Vq}(t_a, \omega)
\right]=\omega {\it Re}\left[ Z(t_a, \omega) \right]$, one obtains
from eq.\ref{Teff} that the effective temperature of the Laponite
solution as a function of the aging time and frequency is:

\begin{equation}
T_{eff}(t_a,\omega)=\pi S_{Z}(t_a,\omega) / 2 k_B {\it Re}\left[
Z(t_a,\omega)\right]
 \label{TLap}
\end{equation}

which is an extension of the Nyquist formula.

\subsection{Experimental results}

\begin{figure}[!ht]
  {\bf (a)} \\
  \centerline{{\epsfysize=0.4\linewidth \epsffile{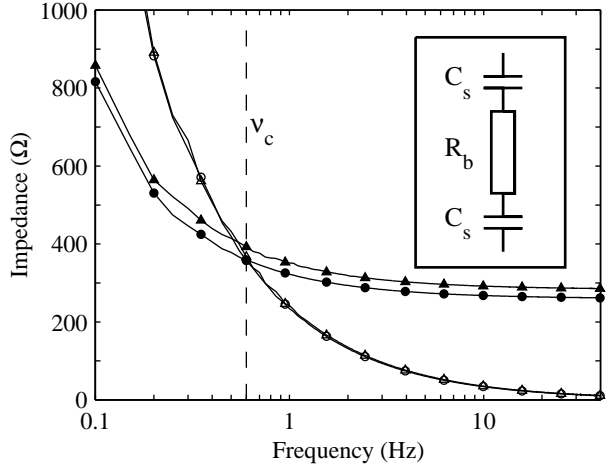}}}
  {\bf (b)} \\
  \centerline{\epsfysize=0.4\linewidth \epsffile{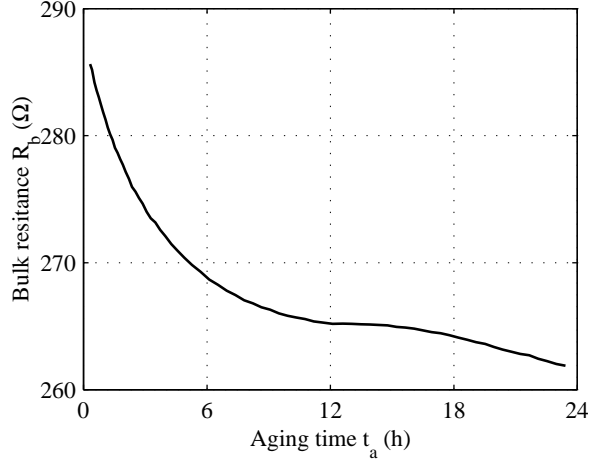}}
  \caption{{\bf The response function} (a) Frequency dependence
  of a sample impedance for 2
different aging time: $t_a=0.3h$, real ($\blacktriangle$) and
imaginary ($\vartriangle$) part; $t_a=24h$, real ($\bullet$) and
imaginary ($\circ$) part. The equivalent circuit for the cell
impedance is shown in the inset: $Z$ is the sum of a resistive
volume $R_b$ and a capacitive interface $C_s$ between the Laponite
solution and the electrodes.  The increase of $Re(Z)$ toward small
frequencies ($\nu<\nu_c$) is due to the dissipative part of the
capacitance (loss tangent $\simeq 0.2$). For $\nu>\nu_c$ the
impedance of the cell is dominated by the bulk resistance $R_b$.
(b) Time evolution of the bulk resistance. This long time
evolution is the signature of the aging of the sol. In spite of
the decreasing mobility of Laponite particles in solution during
the gelation, the electrical conductivity increases.}
\label{response}
\end{figure}

In Fig.\ref{response}(a), we plot the real and imaginary part of
the impedance as a function of the frequency $\nu$, for a typical
experiment. The response of the sample is the sum of 2 effects:
the bulk is purely conductive, the ions of the solution follow the
forcing field, whereas the interfaces between the solution and the
electrodes give mainly a capacitive effect due to the presence of
the Debye layer\cite{hunter}. This behaviour has been validated
using a four-electrode potentiostatic technique \cite{electrochem}
to make sure that the capacitive effect is only due to the
surface. In order to test only bulk properties, the geometry of
the cell is tuned to push the surface contribution to low
frequencies: the cutoff frequency of the equivalent R-C circuit is
less than $0.6 Hz$. The time evolution of the resistance of one of
our sample is plotted in Fig.\ref{response}(b): it is still
decaying in a non trivial way after $24 h$, showing that the
sample has not reached any equilibrium yet. This aging is
consistent with that observed in light scattering experiments
\cite{Kroon}.

\begin{figure}
\begin{center}
        \psfrag{xl}[tc][tc]{\small Frequency $\nu$ ($Hz$)}
        \psfrag{yl}[Bc][Bc]{\small $S_{Z}(t,\nu)$ ($V^2/Hz$)}
        \psfrag{xle}[tc][tc]{\tiny $\omega (t/1h)^{0.5}$ ($rad.s^{-1}$)}
        \psfrag{yle}[Bc][Bc]{\tiny $S_{Z}(t,f)$ ($V^2/Hz$)}
        \psfrag{0.3h}[Br][Br]{\tiny $t=0.3\, h$}
        \psfrag{1.5h}[Br][Br]{\tiny $1.5\, h$}
        \psfrag{2.5h}[Br][Br]{\tiny $2.5\, h$}
        \psfrag{5h}[Br][Br]{\tiny $5\, h$}
        \psfrag{10h}[Br][Br]{\tiny $10\, h$}
        \psfrag{50h}[Br][Br]{\tiny $50\, h$}
        \psfrag{slope}[Bl][Bl]{\tiny slope $\omega^{-3.3}$}
        \psfrag{3}[Br][Br]{ }
        \includegraphics{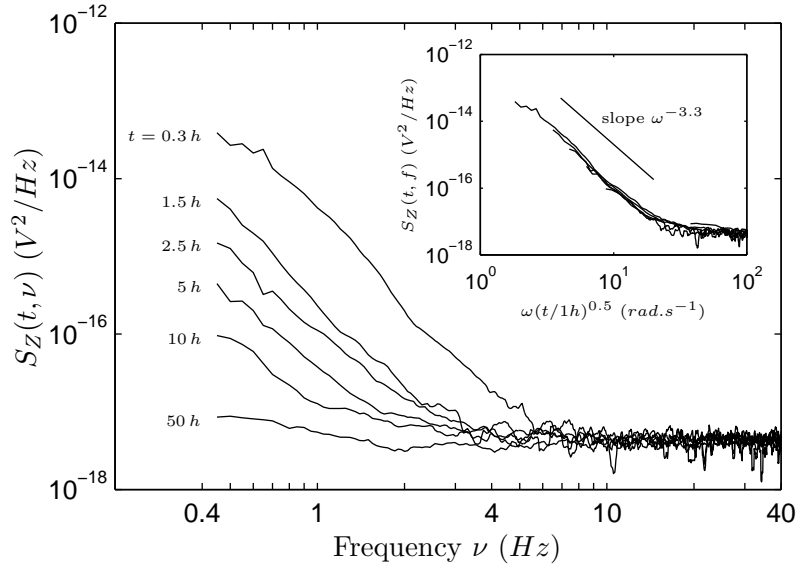}
    \end{center}
\caption{{\bf Fluctuations} Voltage noise density of one sample
for different aging times. The strong increase of $S_Z$ for low
frequencies is quite well fitted by a power law $\omega^\alpha$,
with $\alpha = - 3.3 \pm 0.4$. This effect is a decreasing
function of time, and a good rescaling of the data with a $\omega
t_a^\beta$ law can be achieved as shown in the inset for $\beta =
0.5 \pm 0.1$. }\label{fluctuation}
\end{figure}

\begin{figure}
    \begin{center}
        \psfrag{f}[cc][cc]{\small Frequency $\nu$ ($Hz$)}
        \psfrag{T}[cc][cc]{\small Effective temperature $T_{eff}$ ($K$)}
        \psfrag{t=0.3 h}[Bl][Bl]{\tiny $t=0.3\, h$}
        \psfrag{t=1.5 h}[Bl][Bl]{\tiny $t=1.5\, h$}
        \psfrag{t=2.5 h}[Bl][Bl]{\tiny $t=2.5\, h$}
        \psfrag{t=5 h}[Bl][Bl]{\tiny $t=5\, h$}
        \psfrag{t=10 h}[Bl][Bl]{\tiny $t=10\, h$}
        \psfrag{t=50 h}[Bl][Bl]{\tiny $t=50\, h$}
        \psfrag{TFD}[Bl][Bl]{\tiny TFD}
        \psfrag{7}[Br][Br]{ }
        \includegraphics{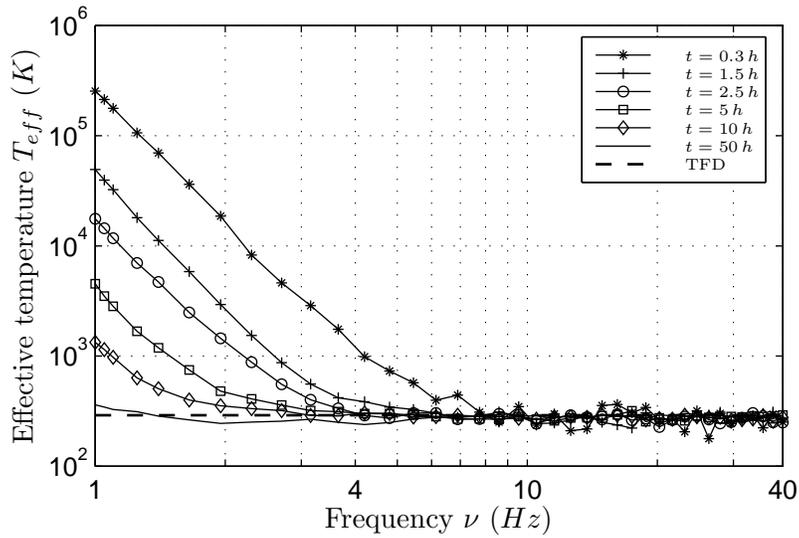}
    \end{center}
\caption{{\bf Effective temperature of Laponite}  Effective
temperature
  as a function of frequency for
different aging times \cite{remarque1}. We restrict the frequency
range to $1 Hz$ to limit the surface contribution to the
measurement. As $S_Z$ in Fig.\ref{fluctuation}, $T_{ef\!f}$
strongly increases and reaches huge values  for low frequencies
and short aging times.  This large violation is observed in
numerical simulations of systems presenting domain growth process.
}\label{temperature}
\end{figure}

\begin{figure}[!h]
    \begin{center}
        \psfrag{f}[cc][cc]{\small Frequency $\nu$ ($Hz$)}
        \psfrag{T}[cc][cc]{\small $T_{ef\!f}$ ($K$)}
        \psfrag{t=0.4 h}[Bl][Bl]{\tiny $t=0.4\, h$}
        \psfrag{t=5 h}[Bl][Bl]{\tiny $t=5\, h$}
        \psfrag{TFD}[Bl][Bl]{\tiny TFD}
        \psfrag{7}[Br][Br]{ }
        \includegraphics{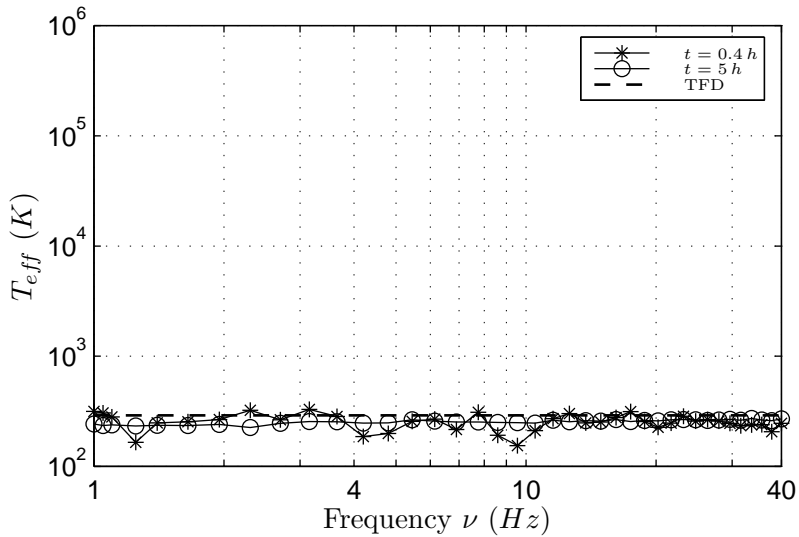}
    \end{center}
\caption{{\bf Effective temperature of an $NaOH$  solution in
water}. The effective temperature is plotted  as a function of
frequency for two different times after the preparation. This
solution has  a {\it p}H close to that of the Laponite, but no
violation is observed in this case  for any aging time.}
\label{fig:electrolyte}
\end{figure}

As the dissipative part of the impedance $Re(Z)$ is weakly time
and frequency dependent, one would expect from the Nyquist formula
that so does the voltage noise density $S_{Z}$. But as shown in
Fig.\ref{fluctuation}, FDR must be strongly violated for the
lowest frequencies and earliest times of our experiment: $S_{Z}$
changes by several orders of magnitude between highest values and
the high frequency tail \cite{1/f}. This violation is clearly
illustrated by the behavior of the effective temperature in
Fig.\ref{temperature}. For long times and high frequencies, the
FDR holds and the measured temperature is the room one ($300K$);
whereas for early times $T_{eff}$ climbs up to $3.10^5K$ at $1Hz$.
Notice that the violation extends to frequencies much larger than
the $0.6Hz$ cutoff separating the volume from the surface effects.
Moreover, the scaling presented in inset of Fig.\ref{fluctuation}
seems to indicate that $T_{eff}$ can be even larger for lower
frequencies and lower aging times. Indeed, we found in all the
tested samples no evidence of a saturation of this effective
temperature in our measurement range.  In order to be sure that
the observed violation is not due to an artifact of the
experimental procedure, we filled the cell with an electrolyte
solution with {\it p}H close  to that of the Laponite sol such
that the electrical impedance of the cell was the same.
Specifically we filled the cell with $NaOH$ solution in water at a
concentration of $10^{-3} \ mol.l^{-1}$. The results of the
measurements of $T_{eff}$ are shown  in fig.\ref{fig:electrolyte}
at two different time after the sample preparation.   In this case
we did not observe any violation of FDR at any time.

\newpage

\subsection{Discussion}

Let us now briefly discussed the results. The observed very large
value of $T_{eff}$ is of course very striking. However the
existence of infinite $T_{eff}$ was predicted \cite{Peliti} and
numerically verified \cite{Barrat} in systems presenting domain
growth process. This is probably the way in which the Laponite
solution makes the transition towards the  colloidal glass state.
The influence of these domains on the electrical conductivity of
Laponite is related to the electrostatic interaction among the
Laponite disks, which is a widely studied and not yet completely
understood problem\cite{Hansen}.  The physical origin of the large
$T_{eff}$ certainly lays in these complex interactions.
 However one may wonder whether the observed $T_{eff}$
is mainly due to bulk effects  or to a conductive phenomenon
produced at the Debye layers on the electrodes. In our opinion the
latter has to be excluded for two reasons. The first one  is that
the violation extends to frequencies much larger than the $0.6Hz$
cutoff   separating the volume from the surface effects.  The
second reason is that complementary measurements, done in cells
with  different geometries, show that for frequency larger than
the cutoff $\nu_c$  the effect of the  Debye layers on the
violation is negligible.

The comparison with the results of the other experiments
\cite{Wendorff,Grigera} is difficult   because these two
experiments are done in different materials and at a single
frequency.  Thus in these two experiments it is impossible to know
the evolution of the violation  both in  frequency  and time.
Furthermore the violation observed in these experiments is of
order one whereas in our experiment it is several order of
magnitudes.
 However our experiment and the one on supercooled fluid
\cite{Grigera} have a common important result, which merits to be
stressed. The violation of FDR at frequency $\omega$ persists for
times which are several order of magnitude larger then the time
$1/\omega$, which is often considered as the characteristic time
 of the violation of FDR at frequency $\omega$.

\section{ Rheological Measurements}

In the previous section we have shown that the FDR is strongly
violated by the electrical properties of Laponite. We want to
understand  whether a violation can be observed in the
measurements of other physical properties. Of course there are no
reasons to assume that $T_{eff}$ is the same for all the variables
but the differences and the analogies found in the evolution of
$T_{eff}$ obtained by the measurements of several variables  may
give new insight on aging theories and on the coupling of the
different variables in an aging material. Therefore we performed
rheological measurements on Laponite and we checked FDR in these
measurements.

\subsection{ Experimental apparatus}
\begin{figure}[!ht]
\centerline{{{\bf a)} \epsfxsize=0.4\linewidth
\epsffile{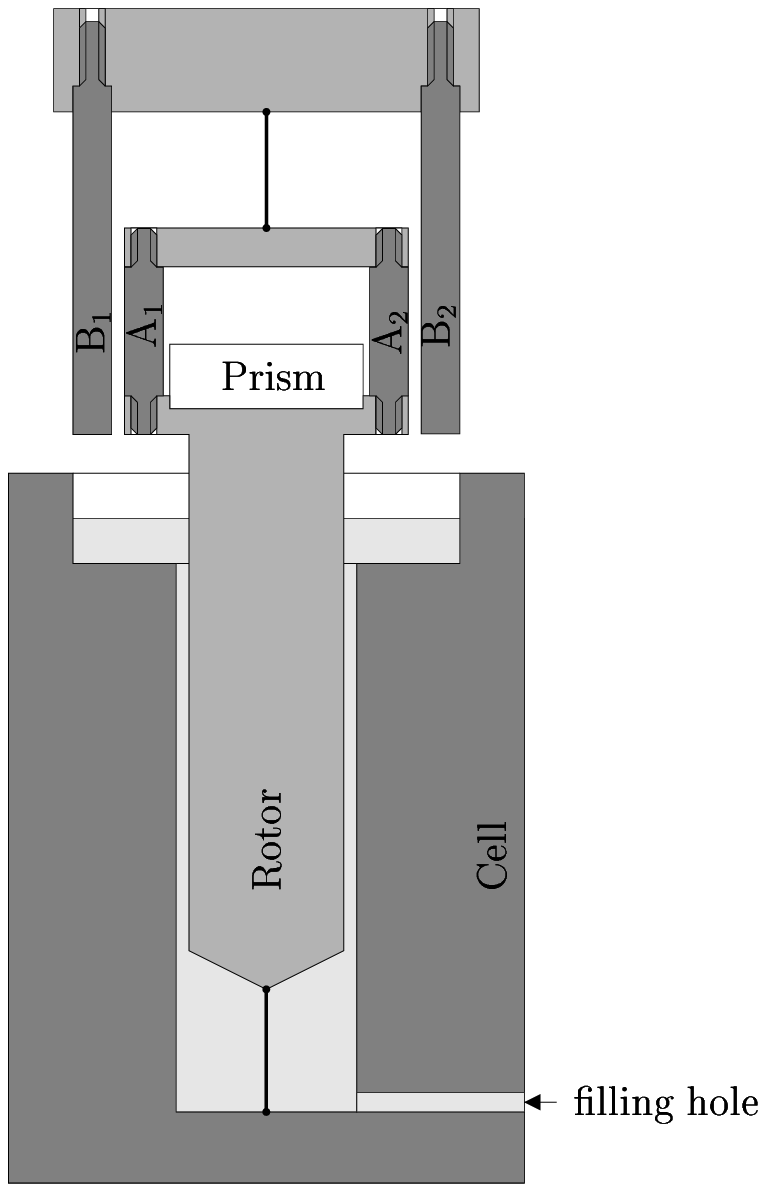}} \hspace{1cm}  {\bf b)}
\raisebox{5cm} {\parbox{7cm} {\epsfxsize=1.0\linewidth
\epsffile{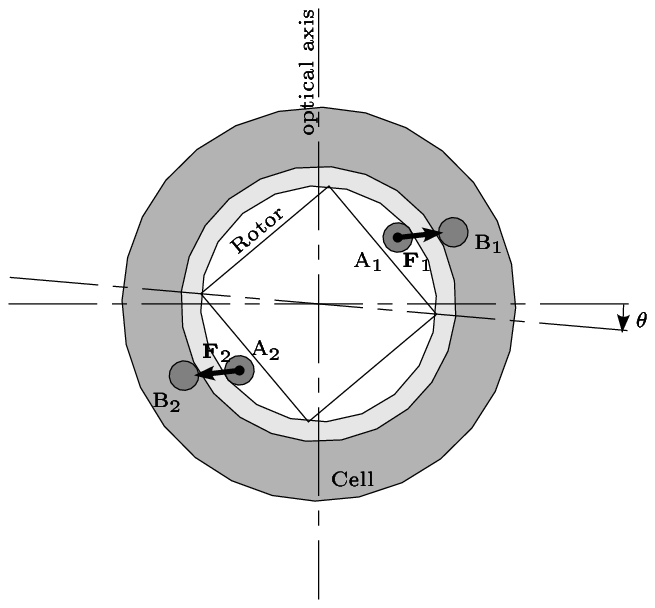}}}}
  \caption{{\bf Rheometer} Vertical (a) and horizontal (b)  cross section
  of the rheometer.
  The torque is applied through the electrostatic interaction of
  electrodes $\text{A}_{1,2}$ and $\text{B}_{1,2}$ when the voltage
  $V_c$ is applied to this capacitor.}
  \label{rheometer}
  \end{figure}

To  achieve this result we built a new rheometer which is
sensitive to thermal fluctuations. The principle of the
rheological measurement is a standard one and is illustrated in
fig.\ref{rheometer}. We describe here only the main features, more
details can be found in ref.\cite{Thesis}. A rotor of diameter
12mm is inserted in a cylindrical cell. The gap of 1mm between the
rotor surface and the cell is filled with the fluid under study.
The rotor is suspended by two steel wires. On the top of the rotor
we fixed an optical prism. This prism is part of a Nomarski
interferometer \cite{Nepoma} which is used to measure the rotation
angle $\theta$ of the rotor. The sensitivity of this system is
better than $10^{-10} rad/\sqrt{Hz}$ corresponding to a torque on
the rotor of about $10^{-13} N\cdot m/\sqrt{Hz}$. An external
torque $\Gamma_{ext}$ can be applied to the rotor by using the
electrostatic interaction of a capacitor (see
fig.\ref{rheometer}). One of the two electrodes of the capacitor
is fixed on the rotor whereas the other is fixed on the cell
walls. When a voltage difference $V_c$ is applied on this
capacitor the attractive force between these two electrodes
produces a torque on the rotor which is equilibrated by the
stiffness of the steel wires.

Let us consider the simple case of a Newtonian fluid of viscosity
$\eta$. The response of this torsion pendulum in Fourier space is
\begin{equation}
\chi_{\theta\Gamma_{ext}}= {\delta \theta \over \Gamma_{ext}}= {1
\over  (k-J\omega^2)- i \alpha \eta \omega}
 \label{rheoresp}
\end{equation}
where $J$ is the  rotor inertia moment, $k$  the steel wires
stiffness and $\alpha$ is  a geometric factor.

In the experiment we have access to two quantities : the angular
position $\theta$ of the rotor, and the voltage $V_c$ driving the
torque. $\Gamma_{ext}$ being quadratic in $V_c$, we add a constant
offset to it in order to linearise this relation. For small
displacements, the response function $\chi_{\theta\Gamma_{ext}}$
is then simply proportional to $\chi_{\theta V_{c}}$. The missing
constant can be found by performing a inertial calibration of the
response : the real part of $1/\chi_{\theta\Gamma_{ext}}$ is a
parabola whose quadratic coefficient is the rotor inertia moment
$J$. $J$ is known with a good precision, and thus can be used to
calibrate the measurement.

\begin{figure}[!h]
    \begin{center}
        \psfrag{xl}[tc][tc]{\footnotesize $\nu$ ($Hz$)}
        \psfrag{ylr}[Bc][Bc]{\footnotesize $\re({\chiup_{\theta \Gamma_{ext}}}^{-1})$ ($Pa.m^3$)}
        \psfrag{yli}[Bc][Bc]{\footnotesize $\im({\chiup_{\theta \Gamma_{ext}}}^{-1})$ ($Pa.m^3$)}
        \includegraphics[width=7cm]{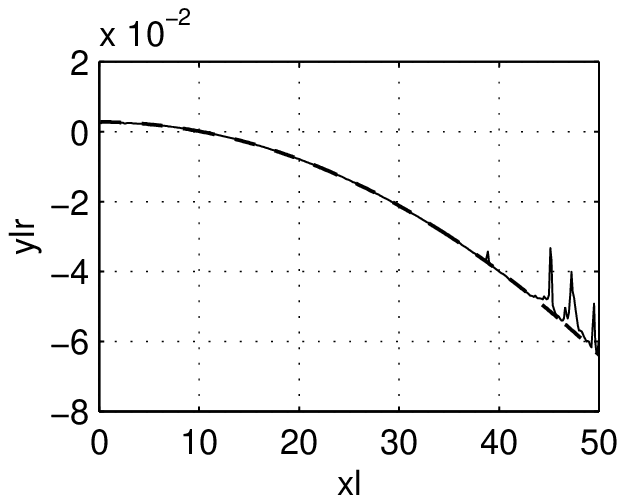}  \hspace{1cm}
        \includegraphics[width=7cm]{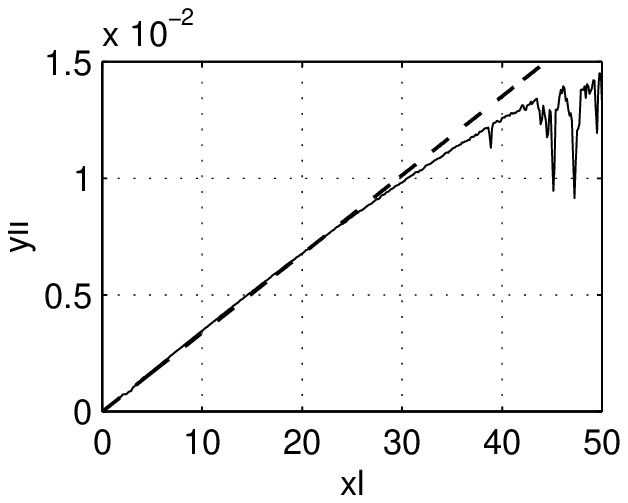}
    \end{center}
  \caption{ Response of the rheometer  to a white noise excitation. The real (a) and
  imaginary (b) part of $1/\chi_{\theta\Gamma_{ext}}$ are plotted as a function
  of frequency. The rheometer
  is filled with a viscous oil with $\eta=2 Pa \ s$. The
  dashed curve in (a) corresponds to a quadratic fit of the
  data, and allows inertial calibration of the measurement (see text
  for details). The dashed line in (b) is a linear fit of low
  frequency data, it is consistent with a newtonian behavior of the
  fluid.}
  \label{fig:rheoresp}
\end{figure}

This rheometer has been tested using a silicon oil with $\eta= 2
Pa \ s$. We first measure the response by using  a white noise
voltage exitation for $V_c$ with an amplitude corresponding to a
torque  $\Gamma_{ext}\simeq 10^{-10} N \ m /\sqrt{Hz}$. The real
and imaginary part of $1/\chi_{\theta\Gamma_{ext}}$ are plotted as
a function of frequency in fig.\ref{fig:rheoresp}(a) and
fig.\ref{fig:rheoresp}(b) respectively. We see that in agreement
with eq.\ref{rheoresp} the real part of
$1/\chi_{\theta\Gamma_{ext}}$ is very well fitted by a parabola
whereas the imaginary part is linear for small $\nu$. The
deviation of the data from the linear behaviour is due to the
non-newtonian character of the silicon oil at higher frequencies.
Once the response is known one can set the external torque to zero
and measure $ S_\theta$, the spectrum of the thermal fluctuations
of $\theta$. In this case the fluctuation dissipation relation is
:
\begin{equation}
 S_\theta= {4 k_B T \over \omega } {\it Im}
(\chi_{\theta\Gamma_{ext}}) \label{rheoFDR}
\end{equation}

\begin{figure}[!h]
    \begin{center}
        \psfrag{3}[tc][tc]{ }
        \psfrag{xl}[tc][tc]{\small $\nu$ ($Hz$)}
        \psfrag{yl}[Bc][Bc]{\small $S_\theta$ ($rad/\sqrt{Hz}$)}
        \includegraphics[width=9cm]{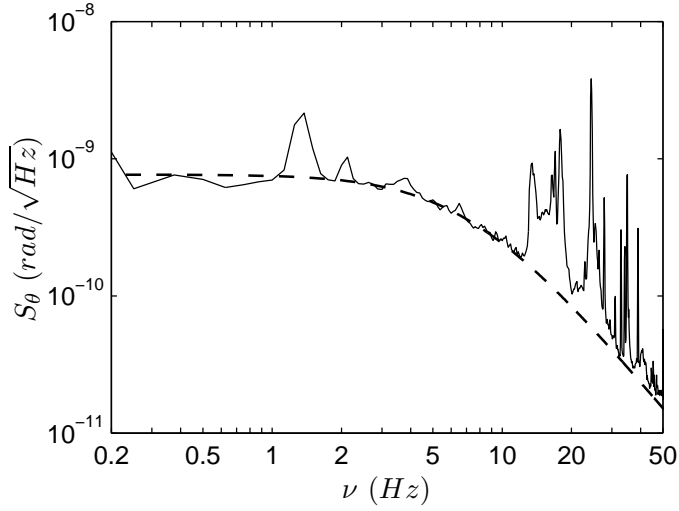}
    \end{center}
  \caption{Spectrum of the thermal fluctuations of $\theta$. The rheometer
  is filled with the same oil used to measure the response reported
  in fig.\ref{fig:rheoresp}.The continuous line is the result of the
  measurement whereas the dashed line is computed inserting in eq.\ref{rheoFDR}
  the measured response of the rheometer and $T=300K$.} \label{fig:rheoFDR}
\end{figure}

By inserting the measured $\chi_{\theta\Gamma_{ext}}$ in
eq.\ref{rheoFDR}, we get an estimation of the fluctuation spectrum
$S_\theta$. The comparison between this computed $S_\theta$ and
the measured one is done in fig.\ref{fig:rheoFDR}. Except for the
existence of the peaks due to environmental noise the agreement is
quite good. Thus the rheometer has enough sensitivity to verify
FDR in viscous fluids.

\bigskip

\subsection{Measure on Laponite}

\begin{figure}[!h]
    \begin{center}
        \psfrag{3}[tc][tc]{ }
        \psfrag{f}[tc][tc]{\small frequency $\nu$ ($Hz$)}
        \psfrag{G1}[Bc][Bc]{\small elastic module $G_1$ ($Pa$)}
        \psfrag{G2}[Bc][Bc]{\small viscous module  $G_2$ ($Pa$)}
        \includegraphics[width=5cm]{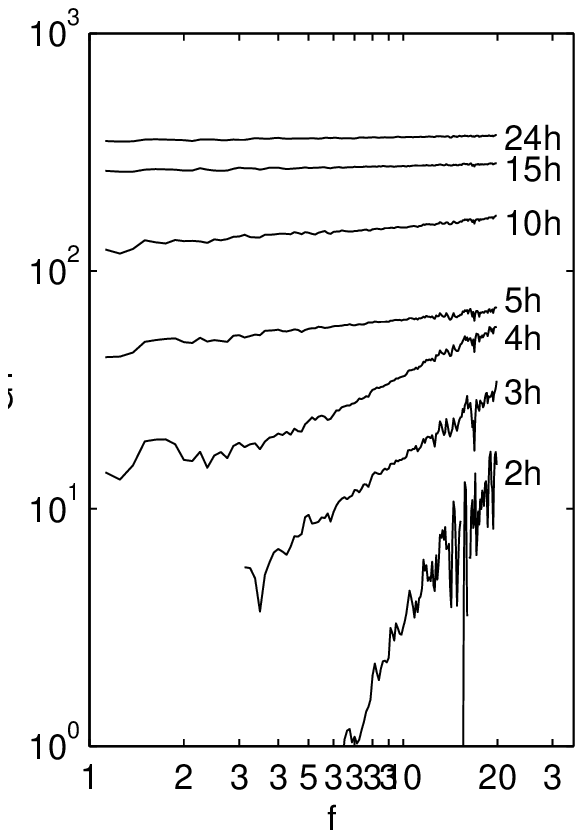}
        \hspace{10mm}
        \includegraphics[width=5cm]{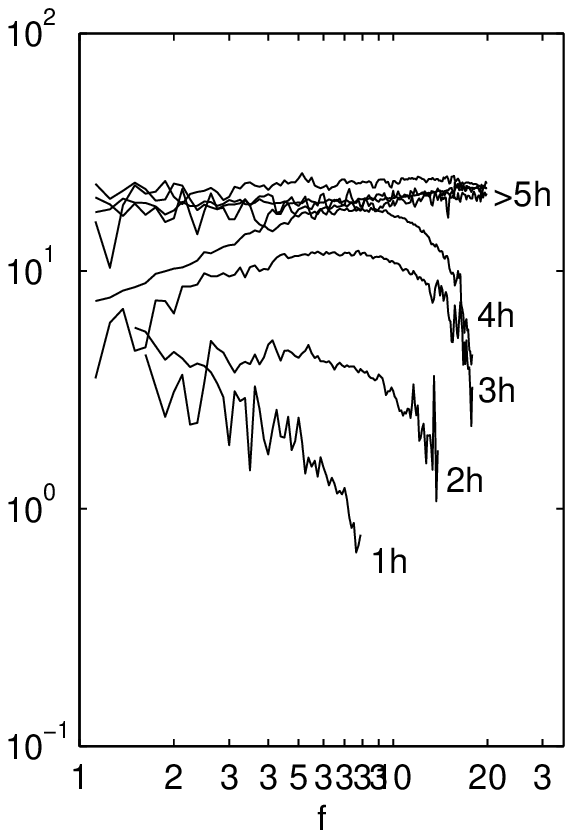}
    \end{center}
  \caption{Aging of the viscoelastic module of Laponite at a $3
  \%$ mass
  concentration in water. The real (a) and
  imaginary (b) part of the viscoelastic module are plotted as a function
  of frequency.
  }
\label{fig:rheolap}
\end{figure}

\begin{figure}[!h]
    \begin{center}
        \vspace{1cm}
        \psfrag{3}[tc][tc]{ }
        \psfrag{300}[cr][cr]{\scriptsize $300 \, K$}
        \psfrag{Teff}[cl][cl]{\scriptsize $T_{ef\!f}$[electric]}
        \psfrag{xl}[tc][tc]{\small $\nu$ ($Hz$)}
        \psfrag{yl}[Bc][Bc]{\small $S_\theta$ ($rad/\sqrt{Hz}$)}
        \includegraphics[width=9cm]{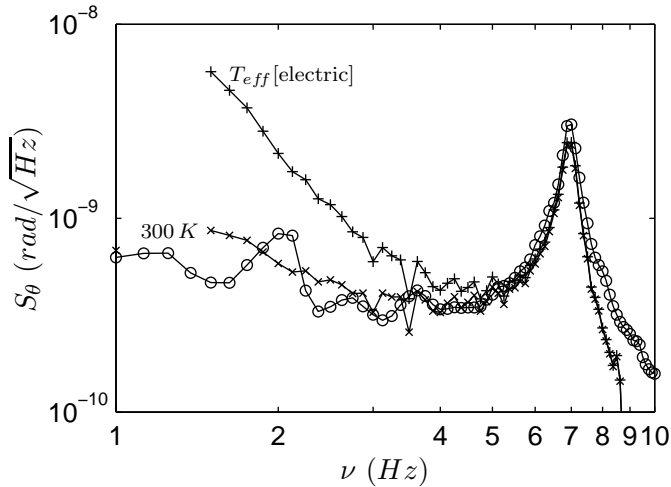}
    \end{center}
  \caption{Thermal fluctuation spectrum of the rheometer filled with Laponite.
  The line with $\circ$ is the  the result of the measurement. The line with $x$
  is the FDT prediction for $T=300K$. The line with $+$ is obtained from eq.\ref{rheoFDR}
  by  inserting in $T$ the values of $T_{eff}$ estimated
 from electric measurements. }
  \label{fig:TFDlapo}
\end{figure}

We have studied the rheological properties of  Laponite at $3 \%$
mass concentration in water. Laponite has been prepared in the way
described in section 2)  for electrical measurements and the
rheometer is mounted inside a container filled with a clean
Nitrogen atmosphere. Moreover, a thin oil layer on top of the
Laponite solution ensures no evaporation can occur. We first
measure the response  at different times $t_w$ after the
preparation. The measured elastic and viscous modules of Laponite
are plotted as a function of frequency in fig.\ref{fig:rheolap}
(a) and (b) respectively. We clearly see that at short $t_w$ the
liquid like solution has only viscous response. As time goes on an
elastic modulus appears and the viscosity  grows of about one
order of magnitude. After 24 hours the solution has a solid like
aspect with a viscoelastic response. The measurement the thermal
fluctuation spectrum $S_\theta$ averaged on the first hour of the
aging processes of Laponite is plotted in  fig.\ref{fig:TFDlapo}.
The peak at $1.5Hz$ has no physical meaning and is due to a
mechanical resonance of the table. Whereas the resonance at $7Hz$
is the resonant frequency of the torsion pendulum of the
rheometer. The measure is compared to the prediction of  the
fluctuation dissipation theorem by inserting in eq.\ref{rheoFDR}
the measured values of the Laponite viscous response and $T=300K$.
We also inserted  in eq.\ref{rheoFDR} the effective temperature
obtained by the electrical measurements averaged on the first
hour. We clearly see that in the case of rheological measurements
no violation of FDR can be detected within experimental errors. If
a violation exists, it  is much smaller than that  observed in the
electric measurements.

\section{Discussion and Conclusion}

We have studied the fluctuation dissipation relations during  the
transition of Laponite from a fluid-like state to a solid-like
colloidal glass using electric and rheological measurements.
Strong aging properties have been observed both in the electrical
and rheological response functions. The behaviour of thermal
fluctuations is instead quite different.  In electric measurements
 a large violation of the dissipation
fluctuation relation is observed at the beginning of the
transition from a fluid like sol to a  glass.  As predicted by the
theory the amplitude and the persistence time of this violation
are decreasing functions of frequency. The observed effective
temperature reaches $ 10^{5} K$ for frequencies smaller than
$1Hz$. This violation involves frequencies which are much larger
than $1/t_w$. Instead in the rheological measurements no
detectable violation of the fluctuation dissipation relation has
been observed. The reasons of this behaviour and  of such a big
difference between the two physical properties  are not
understood. The first striking result is the very high effective
temperature measured in the electrical properties. These very high
effective temperatures has been observed in the numerical
simulation  of aging models presenting domain growth. Thus this
behaviour of electrical properties of laponite could be explained
by the formation of large domains during the transition toward the
colloidal glass state. However one wonders whether the  growth of
these domains does not affect the fluctuations in the rheological
measurements. Another possible explanation of the large  $T_{eff}$
in electric measurements is the dissolution of aggregated
particles and consequently of ions in the Laponite-water solution.
It is conceivable that this effect should not influence the
rheological measurements. The second striking result, which merits
to be discussed, is the fact that in electric measurements the
violation involves frequencies which are much larger than $1/t_w$.
This results agrees with the experiment of ref.\cite{Grigera} but
it is different from what is theoretically predicted and computed
in numerical simulation. This observation leads us to another
possible explanation of the discrepancy between electrical and
rheological properties of Laponite. Indeed the time scale of
relaxation of the fluctuations could not be the same for the two
properties. At the moment we have not a clear and precise answer
to this problem  and  more precise measurements are necessary in
order to understand this large difference between rheological and
electrical measurements.

The results of this paper are certainly  preliminary but they show
that the measurements  of FDR in aging systems can give more
insight to the problem of material aging. Indeed aging has been
always characterized only by the measurements of the response
function. We have shown here that the association of this
measurement to that of the thermal fluctuations may show new and
unattended aspects of an aging material.

\bigskip

\bigskip

{\bf Acknowledgements}

We thank  J.P. Bouchaud, L. Cugliandolo, J. Kurchan, M. Mezard, E.
Vincent, G. Parisi, J.P. Hansen, E. Trizac, and J.-F. Palierne for
useful discussion. We acknowledge F. Vittoz and L. Renaudin for
technical support. This work has been partially supported by the
Programme Th\'ematique of Region Rh\^{o}ne-Alpes.


\begin{thebibliography}{}
\bibitem{1/f} This low frequency noise cannot be confused with the
standard $1/f$ noise observed in many electronic devices. We
recall that the $1/f$ appears only when an external  current
produced by an external potential  goes through the device. In our
cell no external potential is applied.
\bibitem{remarque1} The usual representation of the effective temperature
in simulations is the slope of the response versus correlation
plot, but it is not suited for our experimental data: the system
being almost only dissipative, the response function is close to a
delta distribution, thus FDR is only one point in this
representation.
\end{thebibliography}
\end{document}